\documentclass[%
 aip,
 pop,%
 amsmath,amssymb,
 reprint,%
]{revtex4-1}

\usepackage{graphicx}
\usepackage{dcolumn}
\usepackage{bm}
\usepackage{color}

\begin{document}

\preprint{APS/123-QED}

\title{Exploring the Statistics of Magnetic Reconnection
X-points in Kinetic Particle-in-Cell (PIC) Turbulence}

\author{C. C. Haggerty}
\email{ColbyCH@udel.edu}
\author{T. N. Parashar}
\author{W. H. Matthaeus}
\author{M. A. Shay}
\author{Y. Yang}
\affiliation{Bartol Research Institute, Department of Physics and
Astronomy, University of Delaware, Newark, DE 19716, USA }
\author{M. Wan}
\affiliation{Department of Mechanics and Aerospace Engineering,
	South University of Science and Technology of China, Shenzhen,
	Guangdong 518055, People's Republic of China}
%
\author{P. Wu}
\affiliation{School of Mathematics and Physics, Queens University, Belfast, BT7 1NN, United Kingdom}
\author{S. Servidio}
\affiliation{Dipartimento di Fisica, Universit\'a della Calabria, Cosenza, Italy }

\date{\today}
\begin{abstract}
	Magnetic reconnection is a ubiquitous phenomenon in turbulent plasmas. It is an important part of the turbulent dynamics and heating of space and astrophysical plasmas. We examine the statistics of magnetic reconnection using a quantitative local analysis of the magnetic vector potential, previously used in magnetohydrodynamics simulations, and now generalized to fully kinetic PIC simulations. Different ways of reducing the particle noise for analysis purposes including multiple smoothing techniques are explored. We find that a Fourier filter applied at the Debye scale is an optimal choice for analyzing PIC data. Finlay, we find a broader distribution of normalized reconnection rates compared to the MHD limit with rates as large as 0.5 but with an average of approximately 0.1.
\end{abstract}

\maketitle

\section{Introduction}
Most naturally occurring plasmas are observed (e.g., \citenum{ColemanApJ68,
SaurAA02,MarschLRSP06,Retino07,BrunoLRSP13}) or believed
to be (e.g., \citenum{MacLowApJ99,BanerjeeMNRAS14}) 
in a turbulent state. 
Turbulent plasma dynamics
create strong gradients in the magnetic field, 
leading to conditions
in which the magnetic topology may change 
at kinetic scales. This can produce fast, bursty
outflows associated with magnetic reconnection \cite{Yamada10}. 
Magnetic reconnection
not only mediates the development of turbulence 
but also is very efficient at converting magnetic field energy 
into kinetic energy, both in flows and in 
thermal degrees of freedom. 
It is therefore 
of great importance 
to quantify the role of reconnection in turbulence, an issue 
addressed here 
for collisionless kinetic plasma, which is particularly
relevant for space and astrophysical applications.

The interplay of reconnection and turbulence can be seen as a two-way interaction: 
on the one hand, 
turbulence can establish and sometimes control 
the conditions for reconnection,
and on the other hand,
phenomena associated with 
reconnection, such as exhaust jets, 
can drive turbulence \citep{MatthaeusPFL86,Lapenta08,MatthaeusSSR11}. 
It has been established \citep{ServidioPRL09} 
that MHD turbulence causes magnetic flux tubes
to interact and reconnect, leading to a broad statistical 
distribution of reconnection rates. 
Similarly, external driving of turbulence can induce 
rapid \citep{Lazarian99} large scale reconnection.
\citet{RetinoNature07} showed how 
the dynamics of reconnection and turbulence 
are closely intertwined in Earth's magnetosheath.
In this study we are interested in the 
former problem, reconnection in the midst 
of broadband plasma turbulence, extending the \citet{ServidioPRL09} study
to the case of collisionless plasma.

Large, noise-free, fully kinetic Eulerian Vlasov simulations of turbulence
are at present extremely computationally intensive.
These are essentially out of reach of
present day computers although the hybrid Vlasov variation,
with fluid electrons, is tractable \cite{ServidioJPP15}. 
A less computationally demanding approach to simulating the fully kinetic model is
particle-in-cell (PIC) \cite{BirdsallBook} model, which we employ here. 
As we will see below,
to assess statistical properties of reconnection 
requires identification of physically-correct 
critical points (here, X-points).
For the PIC method this involves not only 
the possibility of 
numerical issues associated with use of finite differences in space,
but also additional 
subtleties connected with finite numbers of macro-particles per cell (or, per 
Debye sphere) \citep{BirdsallBook}. Understanding how this numerical issue can affect the sub-ion scale dynamics of PIC simulations is an important part of accurately simulating collisionless plasma turbulence. Several works have shown that a large number of marco-particles per cell are required to capture these dynamics using an explicit scheme \citep{FranciApJL15,CamporealeCPC16}
In this paper, to achieve reliable and physically correct results,
we study the effects of changing the 
counting statistics as well as the effects of filtering the electromagnetic 
fields. The influence of these variations on different-order statistical quantities
will be considered (i.e the spectrum, the scale dependent kurtosis and the number of X-points). 
Using the appropriate filtering, we analyze the number of X-points generated in a large fully developed decaying turbulence simulation, and thereby 
arrive at a physically reliable assessment of 
the associated reconnection rates in the many flux tube interactions
that occur.

\section{Accuracy of Turbulence in Simulations}
Turbulence involves dynamical activity
that spans broad ranges of 
temporal and length scales, 
a property that makes accurate simulation inherently difficult.
Computing dynamics of large structures
requires long simulation times, while appropriately 
simulating the much faster activity at small dissipation
scales requires fine spatial resolution and well resolved short 
time scales. 
In the midst of these complex dynamics,
understanding the role of magnetic
reconnection at inertial range scales 
is important for
understanding both the topological and the energetic 
features of the turbulent cascade \citep{Matthaeus86, Wan12b,MatthaeusVelli11}. 
To quantify the 
role of reconnection in kinetic turbulence 
a first step must be to locate putative reconnection sites.
In two dimensional turbulence this means finding the magnetic X-type critical points, namely saddle points of the magnetic potential, were the in-plane magnetic field is null.
One needs also to understand whether the 
identified X-points are physical, or if they are 
numerical artifacts. This is not a trivial issue.
In 2D MHD, it has been found 
that the number of X-points generated during turbulence
depended on the magnetic Reynold's number (Number of X-points $\sim \
Re_m^{3/2}$) \citep{WanPP13}, even when care is taken to
control for numerical errors. 
It was also previously shown that
inadequate spatial resolution results 
in the generation of spurious X-points, but that 
the number of critical points converge to a stable 
value when the resolution is 
approximately 3 times smaller than the Kolmogorov dissipation scale
\citep{WanPP10}. 
It is important to note that while the
energy spectra 
in an under-resolved simulation might 
display the correct characteristic
turbulent features, 
the value of different higher order statistics (e.g. the
scale dependent kurtosis or the number of X-points) can be 
incorrectly
calculated due to the grid scale 
Gaussian fluctuations \citep{WanPP10}. 
A sequence of studies 
\citet{Wan09, WanPP10, WanPP13} has led to a reasonable
level of understanding of the generation of
X-points due to both physical and numerical effects for the case of 
MHD turbulence.
However, open questions remain about proliferation of X-points in kinetic plasmas and in observations \citep{GrecoApJ16}.

The above-mentioned background 
makes it clear that even though a simulation might 
qualitatively appear to be properly resolved, 
the smallest scale dynamics, and the topology of the magnetic field,
might not be accurately accounted for. 
An understanding of this issue
is potentially critical for studying kinetic physics 
in turbulent PIC simulations. 
PIC is a Monte-Carlo technique for numerical solution of the Vlasov 
equation that breaks up the distribution function into a large number of ``macro particles'' and then traces their trajectory in both real and velocity space. 
Fields in PIC are susceptible to noise issues as well as entropy conservation violation 
associated with poor counting statistics \citep{BirdsallBook}. 
In a PIC simulation with fixed spatial resolution,
 these statistics are ultimately tied to the number of macro particles per grid cell (ppc). Simulations with an inadequate number of ppc can lead to the gaussianization of real and velocity space gradients on time scales comparable to those of dynamical interest \citep{Montgomery70}. It has previously been shown in turbulent PIC simulations that coherent structures ranging in lengths as small as the electron scale are generated and that the plasma becomes intermittent \citep{Wan12b,KarimabadiPP13}.
The present work is geared towards beginning the process of understanding what 
controls the proliferation of X-points in kinetic simulations. 
However before addressing that question, or even the question of how the 
simulation resolution affects this process, we need to understand if 
and how the number of macro particles affects the higher order 
statistics of smaller scale processes. Here we study this
by using some of the tests outlined in \citet{WanPP10} namely 
the number of X-points and the behavior of the 
scale dependent kurtosis.

\section{Simulation Details}
The simulations were performed using the fully
kinetic 2.5D ($X$, $Y$ in real space, $V_X$, $V_Y$, $V_Z$ in velocity space), electromagnetic PIC code P3D \cite{ZeilerJGR02}. Two sets of
simulations are studied. The first set is the Orszag-Tang vortex
(OTV) setup\cite{OrszagJFM79, DahlburgPFB89, ParasharPP09, VasquezApJ12}, 
performed in a doubly periodic domain of $(10.24 d_i)^2$ with a
grid spacing and time step of $\Delta x = .02 d_i$ and $\Delta t = .0015 \Omega_{ci}^{-1}$
respectively. The simulations uses an artificial mass ratio of $m_i/m_e = 25$ and speed of light $c/c_{A0} = 20$ where $c_{A0}$ is the Aflv\'en speed based on a reference density and magnetic field of 1 in simulation units. The simulation begins with a uniform
density, a mean out-of-plane magnetic field of 1, 
and in-plane
magnetic and velocity fluctuation of r.m.s. strength 0.2. 
The ion and electron temperatures
are initially uniform with $T_i = T_e = .3 m_ic_{A0}^2$. Five simulations were performed,
varying the number of particles per cell (ppc =12, 50, 200, 800, 3200), 
but keeping all other parameters fixed. 
The simulations were performed to times 
just after the peak of dissipation (peak in $|J_z^2|$) at $15 \Omega_{ci}^{-1}$. This is slightly less than $2 \tau_{nl}$ where $\tau_{nl}$ is the large scale nonlinear
time (eddy turnover time) of the system. 
This system at this time is ideal to test the reconnection site
finding algorithm, as the number of physical reconnection sites can be visually identified and counted.

The second simulation is $(102.4 d_i)^2$ in size with a grid spacing
of $\Delta x = 0.0125 d_i$ and $\Delta t = 0.0025 \Omega_{ci}^{-1}$ and $c/c_{A0}=30$.
The simulation was initialized with an MHD like initial condition with ``Alfv\'enic perturbations'' in the in-plane $B$ and $V$ with an initially specified spectrum. This
simulation was preformed to study von-K\'arman energy decay in kinetic plasmas and more details about the simulation can be found in \citet{WuPRL13}. Three different time snapshots of the simulation at times $t= 206.25, 250.0, 292.5\ \Omega_{ci}^{-1}$ (where $\Omega_{ci}^{-1}$ is based on the mean out-of-plane magnetic field of 5) are used in this study for better statistics. Details for all of the simulations are presented in Table~\ref{tab:table0} along with the number of macro-particles in a Debye circle $N_{\lambda_D}$ where $N_{\lambda_D} = \pi \lambda_D^2\times \rm{ppc}/(\Delta x)^2$
\begin{table}
	\caption{Parameters for different turbulent simulations. The simulation length in $d_i$ ($l_x$, $l_y$), the grid spacing in $d_i$ ($\Delta x$), the time step ($\Omega_{ci}^{-1}$ based on a uniform magnetic field of $B_0 = 1$), the speed of light $\frac{c}{c_A}$, the ion to electron mass ratio ($\frac{m_i}{m_e}$), the number of particles per cell (ppc) and the number of particles per Debye circle ($N_{\lambda_D} = \pi \lambda_D^2\times \rm{ppc}/(\Delta x)^2$).}
\label{tab:table0}
	\begin{ruledtabular}
		\begin{tabular}{c c c c c c c c c}
			Name & $l_x = l_y$ & $\Delta x$ & $\Delta t$ & $\frac{c}{c_A}$ & $\frac{m_i}{m_e}$ & ppc & $N_{\lambda_D}$ \\
			\hline
			OTV1 & 10.24 & .02 & .0015 & 20 & 25 & 12 & 71\\
			OTV2 & 10.24 & .02 & .0015 & 20 & 25 & 50 & 295\\
			OTV3 & 10.24 & .02 & .0015 & 20 & 25 & 200 & 1178\\
			OTV4 & 10.24 & .02 & .0015 & 20 & 25 & 800 & 4712\\
			OTV5 & 10.24 & .02 & .0015 & 20 & 25 & 3200 & 18850\\
			PWu1 & 102.4 & .0125 & .0025 & 30 & 25 & 400 & 11170\\
		\end{tabular}
	\end{ruledtabular}
\end{table}

\section{Analysis Methods}
\subsection{Noise reduction}
As alluded to above, any attempt to identify and tabulate 
X-points, active reconnection sites, and associated reconnection
rates in two dimensional turbulence models must confront 
both physical and numerical issues, and importantly, 
must distinguish between 
them.
In this regard a useful observation is 
that the onset of numerical resolution issues in MHD are signified 
by the appearance of phase errors and ``gaussianization'' of noisy 
fluctuations at the smallest scales \cite{ServidioPP10,Wan10}.
Apart from proper resolution to determine the physical 
reconnection rate, an added complication
is that at high magnetic Reynolds numbers, the number of X-points
increases due to {\it physical} effects \cite{WanPP13}, as the most 
intense current sheets themselves become turbulent 
and unstable at high local Reynolds numbers.

For the OTV configuration in MHD and at moderate large 
scale Reynolds numbers (say, $R_m < 1000$), one would not expect 
physical secondary islands to form dynamically, using the 
well-known empirical criterion (threshold of $R_m \sim 10,000$)
suggested by \citet{Biskamp86}. As far as we are aware, for kinetic 
plasma a similar threshold is obtained (see, e.g. 
\citet{Daughton09}) with the system size 
determining the effective Reynolds number \cite{MatthaeusPRL05}. 

Carrying the above ideas over to the present case
of kinetic simulation,
we expect that 
the number of reconnection sites in a moderate size (Reynolds number) 
OTV simulation at the peak of current
to be $\sim 4$. 
However if a snapshot is taken from a PIC simulation 
at this time and analyzed directly, 
the number of X-points may be found to be as large as 
$\sim 10^4$, depending on parameters. 
It transpires, as we will show below,
that particle noise has to be removed before the physical X-points 
can be unmasked from the ``noise X-points''. 
For the results discussed here 
we will consider three different approaches to 
to reducing the
particle noise:

\begin{itemize}
	\item {\em Increasing the number of particles per cell:} Ideally
	      PIC simulations should be run with as many particles per cell (ppc) as
	      possible. However, computational limits often 
restrict the choice to a few
	      hundred for typical simulation. The particle noise reduces as $1/\sqrt{N}$
	      where $N$ is the number of ppc. In this study we varied
	      ppc from 12 to 3200.

	\item {\em Time averaging:} A technique for reducing particle noise in
	      PIC simulation output data is to time average the data over time scales
	      much shorter than those of dynamical interest. This averaging can reduce plasma oscillations associated with finite particle number charge imbalances.
          In this study
	      we employed time averaging of the data over a period of 
$3 \Omega_{pe}^{-1}$, which is roughly
	      200 times smaller than the typical nonlinear time (for the $(102.4 d_i)^2$ simulation.

	\item {\em Gaussian/ Fourier Spatial Filtering:} A standard technique used in
	      studying the scale to scale transfer of energy is real space
               \cite{GermanoJFM92,EyinkPhysD05}
or sharp Fourier space filtering \cite{AluieEyinkPoP09,WanPRL12}
       In this work we will employ a Gaussian filter as for our real space filtering, defined as:
	      \begin{equation}
g(B_{i,j}) = \sum_{k,l}\frac{B_{k,l}}{\sqrt{2\pi}\lambda_d/\Delta x}e^{-\frac{(i-k)^2 + (j-l)^2}{2(\lambda_d/\Delta x)^2}}
	      \end{equation}
	      where each i, j grid point is smoothed by neighbors with $k$ and $l$ truncated
	      at $\pm 4$. For Fourier filtering, we apply a sharp cutoff
	      in Fourier space at a scale of interest $k_s$. For both Gaussian and
	      Fourier filtering we applied the filters at the scale $\lambda_d$
	      for each simulation.
\end{itemize}

We apply these techniques 
in concert with the development of the 
specific diagnostics that will be used to 
analyze reconnection properties. Subsequently we 
turn towards the physical properties revealed by the statistics of
reconnection. 
It should be noted that out of the three noise reduction procedures 
mentioned above, only the increase in number of particles per cell 
will reduce the effects of noise in the time evolution of the simulation. 
Time averaging and Gaussian/ Fourier filter are post-processing 
procedures that do not remove the problems
associated with particle noise during the time evolution of the
simulation. 

\subsection{Identifying Critical Points in 2D PIC Turbulence}
To identify the critical points in 2.5D simulations, we follow the
procedure devised by \citet{ServidioPRL09, ServidioPP10}. We examine the
1st and 2nd derivatives of the magnetic vector potential.
Magnetic null points are identified as zeros 
of the 1st derivatives. At the null points
we consider the sign of
the product of the eigenvalues of the second derivative (Hessian) Matrix,
$M_{i,j} = \frac{\partial^2 a}{\partial x \partial y}$.  Two 
negative eigenvalues indicate a maximum of the magnetic vector potential,
 and  two positive eigenvalues indicate a minimum.
Null points where the product of the eigenvalues is negative  
are identified as saddle points, and possible physical X-points. 

\section{Effects of time averaging and spatial filtering}

\begin{figure}
	\includegraphics[width=3.2in]{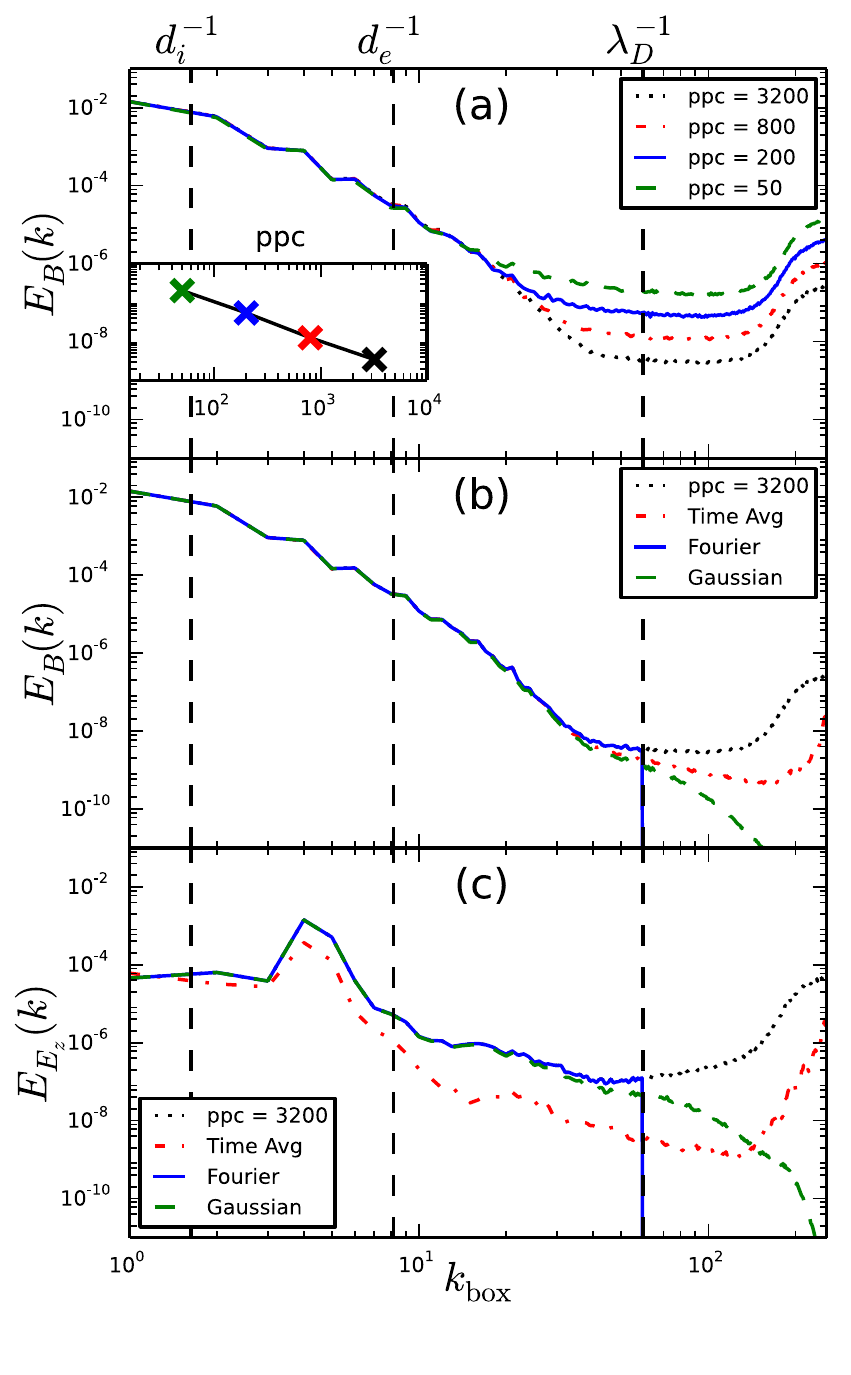}
	\caption{Omnidirectional spectrum of the magnetic field (a + b)  and
		out-of-plane electric field (c) for OTV simulations. Cases are shown with different particles per cell (a) and employing time averaging, Gaussian filtering and Fourier filtering for the 3200 ppc simulation (b + c). The subplot inside (a) shows the value of the magnetic spectrum
		at the Debye length ($E_B(k\lambda_d = 1)$) as a function of particles
		per cell. The slope of the best fit line is approximately -1 which implies that the
		noise floor is inversely proportional to the number of particles per cell). The vertical dashed lines correspond to the wave number 
        of 3 major lengths scales: the ion inertial length, electron internal length and the Debye length from left to right respectively}
	\label{spectrum}
\end{figure}

We start by comparing the results of smoothing on turbulent statistics. 
First we examine power spectra of electromagnetic fields. Fig~
\ref{spectrum} shows the omnidirectional spectra of the magnetic field
and out-of-plane electric field for the OTV simulations. The top panel shows
the magnetic field spectra for varying number of particles per cell. The
result, as expected, is that the particle noise goes down with increasing
number of particles, and this is reflected in the spectra. 
The discrepancies in magnetic spectrum for different ppc simulations can be seen at scales as large as $kd_e \sim 2$. The insert in the top panel shows the power
in magnetic field at the Debye scale as a function of ppc. As the noise in a variable is
expected to go down as $\sqrt{N}$ with number of particles, the energy
is expected to go down linearly. This is indeed the case because the slope of the line in the figure is -1.

The lower two panels of Fig. \ref{spectrum} show the magnetic
and out-of-plane 
electric field ($E_z$) spectra for the largest
ppc run for time averaging as well Gaussian and Fourier filtering.
Time averaging the magnetic field only changes spectrum very close to the Debye
scale, reducing the noise by a factor of a few. The Gaussian filtered data also matches the unprocessed data up to the Debye scale. The Fourier filter by definition matches identically until its sharp cutoff at Debye scale. 

\begin{figure}
	\includegraphics[width=3.2in]{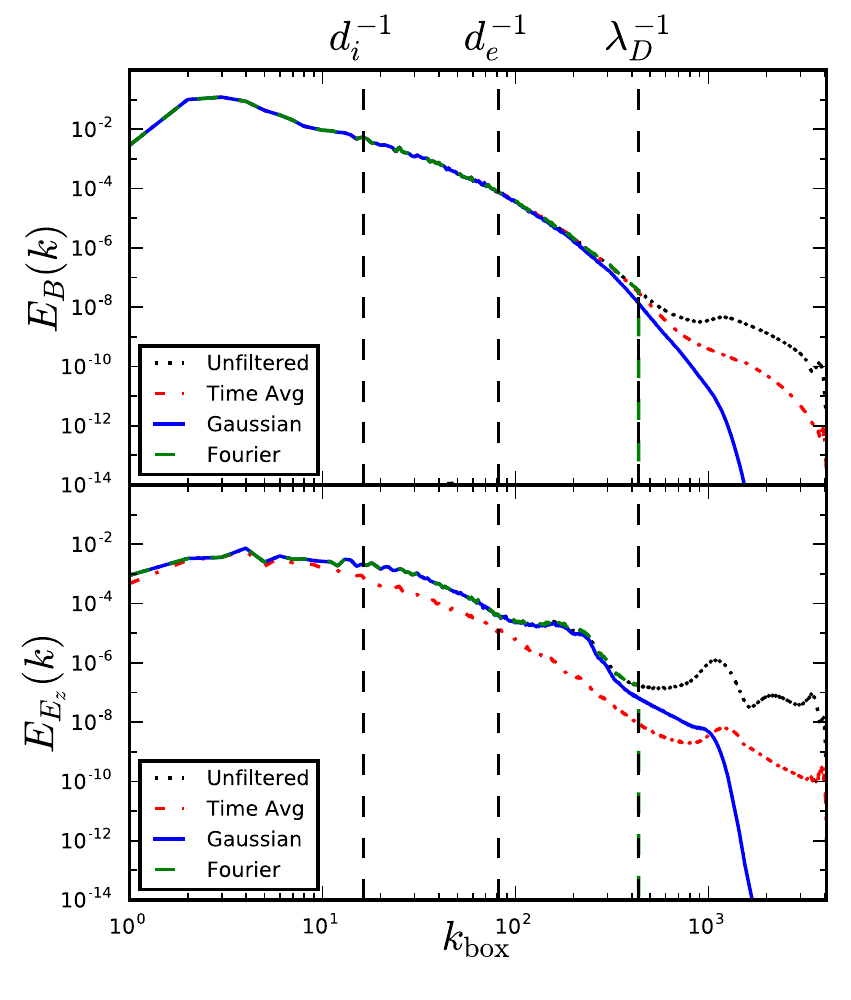}
	\caption{Omnidirectional spectrum of the magnetic field
    (a) and out-of-plane electric field (b) respectively for time averaged, Gaussian filtered, Fourier filtered and unfiltered fields for the large turbulent simulation. The three vertical dashed lines correspond to the wave numbers associated with the Debye length, the electron inertial length and the ion inertial length.}
	\label{spectrum_big} 
\end{figure}

The most drastic effect of time averaging is
visible on $E_z$. Time averaging on a few plasma frequency
time scales reduces spectral power even at scales $\sim d_i$. This implies
that the time averaging has drastic effect on electric fields.
The adverse effect of time averaging on the electric field is also
apparent in the spectra of the large PIC run as shown in Fig.
\ref{spectrum_big}. Broadband reduction of the electric field 
by time averaging can suppress the computed estimates of 
scale to scale transfer of energy
\cite{YangPoP17SUB}. An understanding of whether this is
a consequence of the numerical method, or if it 
relates to high frequency physics 
remains to be investigated. 

\begin{figure}
   \includegraphics[width=3.2in]{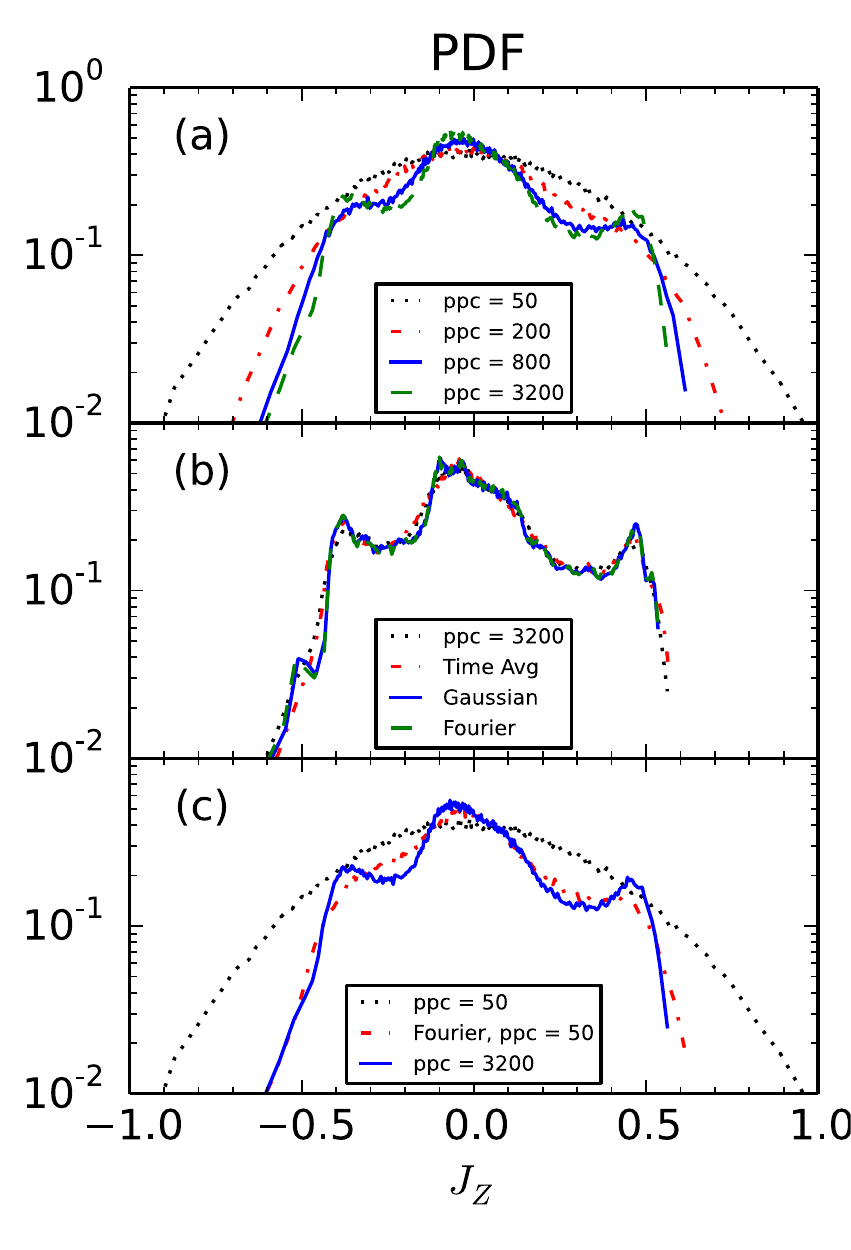}
   \caption{PDFs of the out-of-plane current density ($J_z$) for different ppc OTV simulations. In (a) we plot the PDFs of $J_z$ with different ppc values ranging between 50 and 3200. (b) shows the PDFs of the time averaged, Gaussian filtered, Fourier filtered and unfiltered $J_z$ for the 3200 simulation. Panel (c) shows the effect of Fourier filter on the PDF of the 50 ppc simulation and compares it  with the unfiltered 50 and 3200 ppc simulation PDFs.}
\label{jzpdf}
\end{figure}

To look at the effects of smoothing on statistics of turbulent
fluctuations, we plot probability distribution functions (PDFs) for
the out-of-plane electric current density ($J_z$) in the OTV simulations. 
Fig~\ref{jzpdf} shows
the PDFs for various ppc and smoothing techniques. The smallest ppc simulations are dominated
by random fluctuations associated with poor counting statistics and the 
corresponding PDF of current density
resembles a Gaussian PDF. 
However, when the ppc = 50 simulation is smoothed with the Fourier filter, the shape of 
the current density PDF converges towards the ppc = 3200 case (see Fig.~\ref{jzpdf}c).

Next we turn our attention to the scale dependent kurtosis (SDK), one
of the common measures of intermittency in a turbulent 
system \citep{FrischBook}. The scale dependent kurtosis is defined as:
\begin{equation}\label{eq:sdk}
\rm{SDK}(r) = \frac{\left < \delta B_x(r) ^4 \right >}{\left < \delta B_x(r)^2\right>^2}
\end{equation}
where the angled brackets denote a spatial average
 and $\delta B_x(r)$ is the increment of the magnetic field in the x-direction, 
defined as $\delta B_x(r) = B_x(x+r) - B_x(x)$. 
In principle the scale dependent kurtosis can be calculated for the increment of any field and 
any vector component;
however in this work we elect to only present the scale dependent kurtosis of $B_x$.
It should be
kept in mind that the OTV simulation, being relatively small in size, has
a very small effective Reynolds number and hence can not have large kurtosis
\cite{ParasharApJ15}. Moreover, the large scale inhomogeneity of OTV
makes the kurtosis drop below 3 at larger scales. However, the scale dependent kurtosis can still
be computed and its convergence to a stable value for different noise
reduction techniques can be studied.

\begin{figure}
	\includegraphics[width=3.2in]{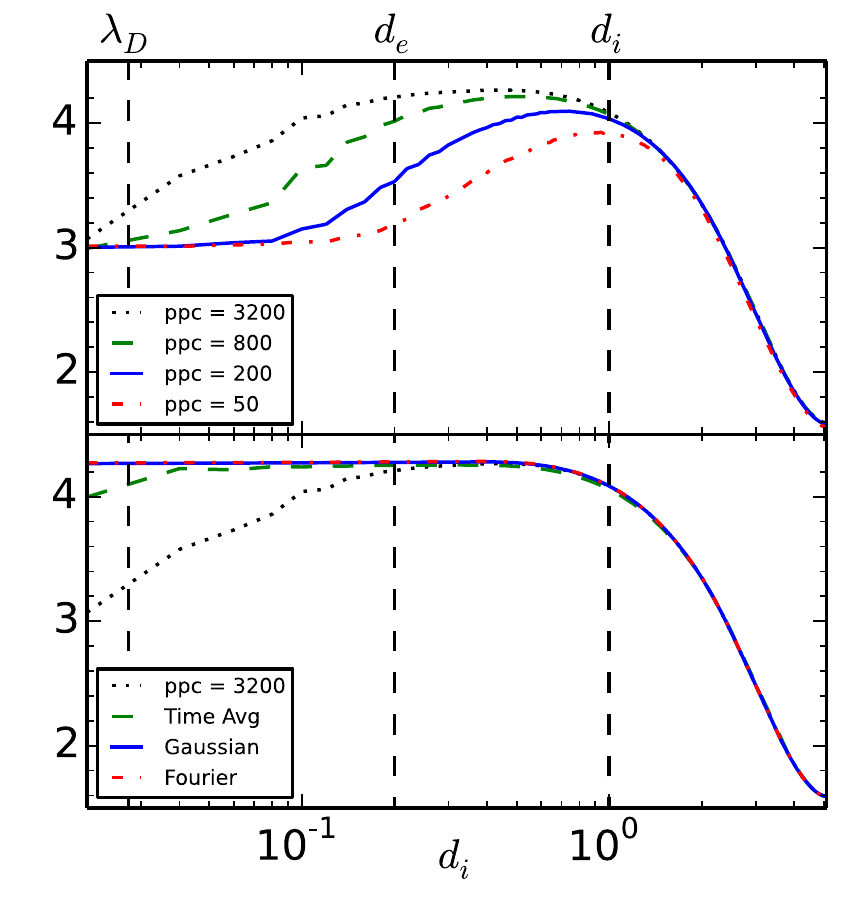}
	\caption{Scale Dependent Kurtosis of the magnetic field ($B_x$) for the OTV simulations with different numbers of ppc (a) and with the field time averaged, Gaussian filtered, Fourier filtered and unfiltered for the ppc = 3200 simulation. The three vertical dashed lines correspond to the Debye length, the electron inertial length and the ion inertial length}
    \label{sdk}
\end{figure}

The top panel of Fig. \ref{sdk} shows the scale dependent kurtosis of $\delta B_x$ for different
ppc cases of the OTV simulation. At large scales the scale dependent kurtosis matches for all
simulations. However, at smaller scales, higher ppc simulations have significantly
larger kurtosis. Particle noise in lower ppc simulations 
evidently randomizes  (gaussianizes) 
the smaller scale structures, decreasing the kurtosis at smaller scales. 
The bottom panel of Fig~\ref{sdk} shows the scale dependent kurtosis of $\delta B_x$ for the ppc 3200 OTV run for different
smoothing techniques. The problem of small scale gaussianization are
alleviated by almost all of the processing techniques and the kurtosis
saturates to a constant value at the smallest scales \cite{WanPRL15}.

\begin{figure}
	\includegraphics[width=3.2in]{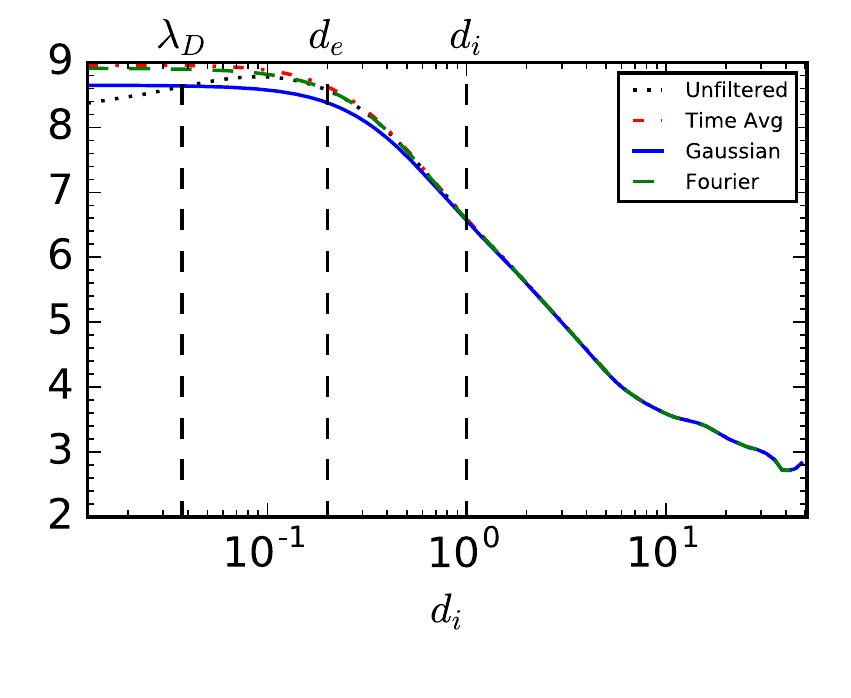}
	\caption{ The scale dependent kurtosis of $B_x$ for the 102.4 x 102.4 $d_i$
	simulation with different smoothing applied.}
\label{sdk_big}
\end{figure}

The scale dependent kurtosis for the larger PIC run (Pwu1 in Table~\ref{tab:table0}), however,
tells a slightly different story. This simulation's larger size allows a
greater separation between the energy containing scales and the
``dissipative'' scales (i.e. kinetic ion and electron scales) and thus
has a larger effective Reynolds number. The larger Reynolds number allows the generation of stronger small scale coherent structures, and thus the energy cascade of this simulation more closely resembles the energy transfer in the turbulent MHD limit \citep{ParasharApJ15}. Fig~\ref{sdk_big} shows the scale dependent kurtosis for $\delta B_x$ for different averaging/filtering techniques. In all cases, scale dependent kurtosis matches very well down to $\sim 0.5d_i$, at which point the Gaussian filtered data starts to diverge from the
other curves. Unfiltered, Fourier filtered and time averaged data sets
match with each other down to scales $\sim 0.5 d_e$. This implies that
a Gaussian filter applied to field data cannot capture the smallest scale intermittent
structures as well as a Fourier filter.

\begin{figure}
	\includegraphics[width=3.2in]{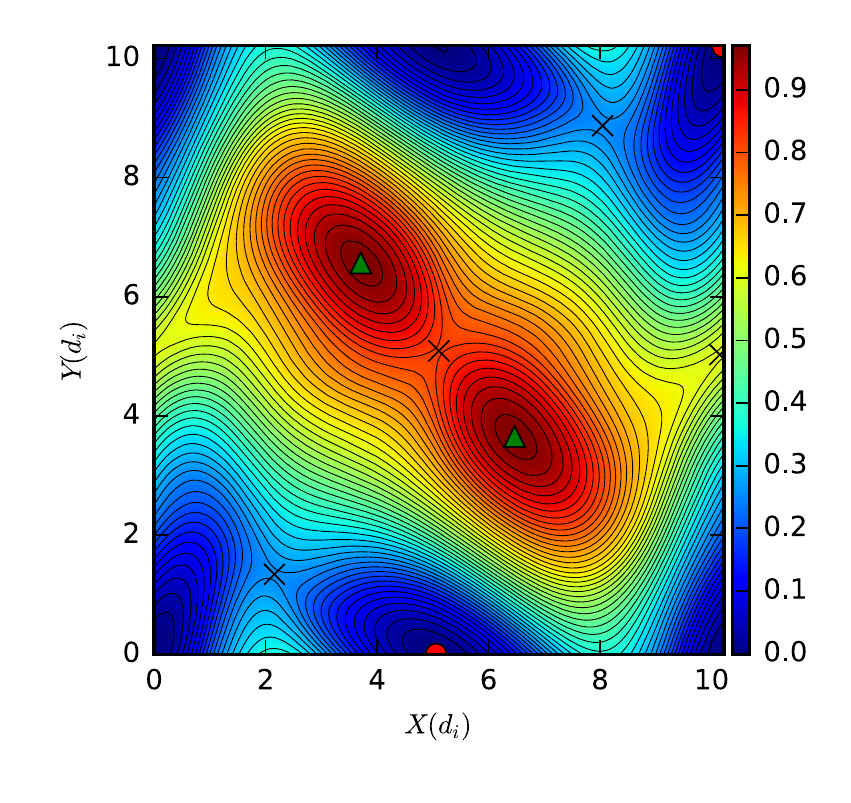}
	\caption{Overview of the 3200 particles per cell Orszag-Tang Simulation 
		simulation where we plot the Fourier filtered magnetic
		vector potential $a$ and with `x' denoting the location of identified
		critical saddle points and o's and triangles denoting minimum and maximum
        respectively. From this overview it is clear that there should only be 4
        X-points, and 4 max/minimums}
\label{otv_psi} 
\end{figure}

\begin{figure}
	\includegraphics[width=3.2in]{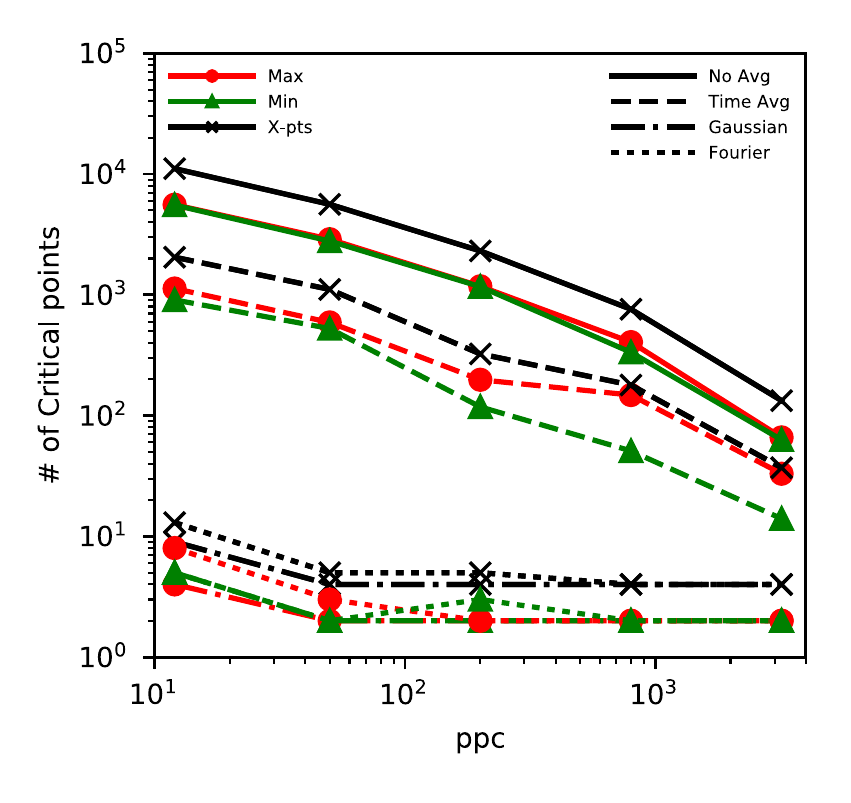}
	\caption{Number of critical points vs particles per cell (ppc) in Orszag-Tang simulations with different field filtering included. Red, green, and black lines correspond to the number of maximum, minimum, and saddle critical points identified in each simulation respectively. The solid line is for the critical points identified from the unfiltered magnetic fields, the dashed lines for the time averaged fields, the dot dashed line for the Gaussian filtered fields, and the dotted line for Fourier filtered fields.}
\label{crit_pts_ppc}
\end{figure}

Finally we study the number of identified X-points for simulations
with different numbers of ppc as well as the effects of time averaging, Gaussian filtering and Fourier filtering.
For this purpose, we choose to work with the OTV near the
peak of mean square electric current density. 
At this time the number of reconnection sites can be visually counted to
be 4 as shown in Fig. \ref{otv_psi}. If the critical point finding
algorithm is applied directly to the unprocessed simulation data, the
noise introduces artificially large number of minima and maxima. Fig~
\ref{crit_pts_ppc} shows the number of critical points as a function
of number of particles per cell. For very small number of particles
the number of critical points is $\sim 10^4$. The number of critical
points decreases as a power law with increasing ppc. However, the
number of particles required to achieve convergence to the 
physical number
of critical points is very large ($> 10^5$). 
Time averaging brings down the number of X-points
significantly (an order of magnitude less) but the number still is 
significantly
large and these cases 
also follow a power-law decrease with increasing ppc. 
On the
other hand, Fourier filter and Gaussian filter (both at the Debye scale)
remove the particle counting 
noise to reveal the physical critical points even for rather small ppc. 
Even
for ppc = 12 the number of identified critical points is less
than a factor of three too large.

Combining the results of analysis of turbulence quantities, and 
the results of critical point finding, we can
conclude the following:
\begin{itemize}
	\item Ideally one would run a simulation with large number of
	particles per cell. However the number of particles required to
	reduce noise significantly and hence capture physical reconnection
	sites is prohibitively large. This would require more computational
    time and so restricts the size of the simulation.

	\item Time averaging, although a common and simple technique
	fails to capture the physical reconnection sites and also adversely
	affects the electric fields for analysis purposes.

	\item Gaussian filtering appears to capture the physical
	reconnection sites and spectra very well but has a slight negative
	effect on the scale dependent kurtosis.

	\item Fourier filtering appears to reduce the noise effects while minimally interfering with the physical effects
discussed above.

\end{itemize}
Based on the above considerations, we conclude that the optimum method to
analyze PIC simulations for reconnection
studies is via the application of a Fourier filter as was done by \citet{Wan12b}. We now identify the
reconnection sites in the large PIC simulation to study their statistics.

\section{Reconnection in Turbulence}
To examine the statistics of reconnection in turbulence, we analyze
the larger 2.5D turbulent PIC simulation again carried out with P3D (PWu1 in Table~\ref{tab:table0}. 
We collect
statistics from three different times in the same simulation ($t = 206.25$, $250.0$ and $292.5\,\Omega_{ci}^{-1}$) For each time sample 
we apply a Fourier
filter to the magnetic field data for $k\lambda_d > 1$ ($\lambda_d =
.0375 d_i$). The results from the topological analysis of the filtered
data for these three sets of output can be found in Table~\ref{tab:table2}.

\begin{figure}
	\includegraphics[width=3.2in]{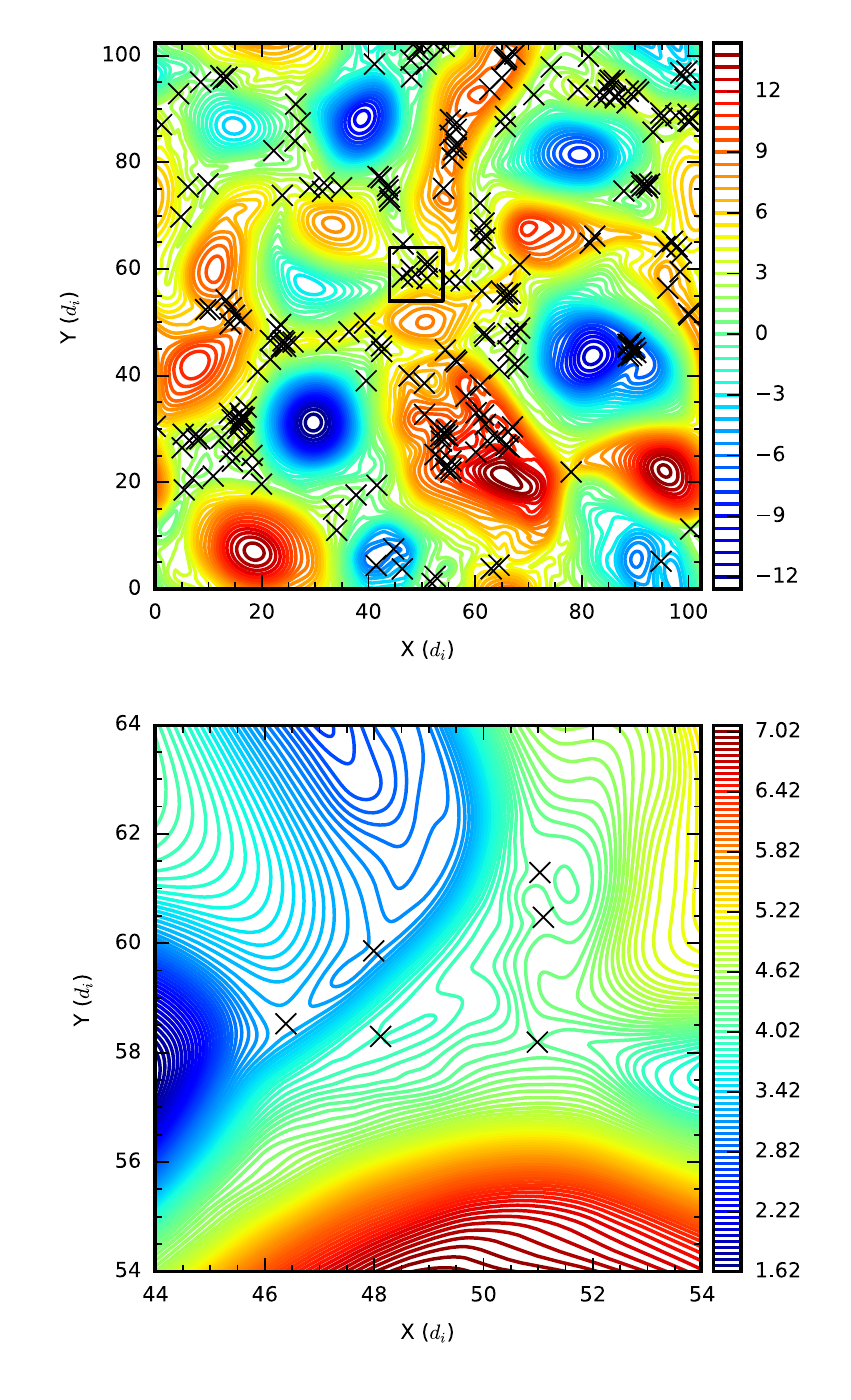}
	\caption{(a) Overview of the large turbulent
		simulation where we plot colored contours of the Fourier filtered magnetic
		vector potential $a$ and with `x' denoting the location of identified
		critical saddle points. (b) is an enlarged subsection from (a) as
        identified by the black box. It is clear from this figure that these
		topological structures exist at different scales throughout the entire
		simulations, and clearly correspond to apparent coherent structures.}
\label{big_psi}
\end{figure}

\begin{table}
	\caption{Number of critical points at each different time.}
	\label{tab:table2}
	\begin{ruledtabular}
		\begin{tabular}{c c c c c}
			Time $\Omega_{ci}^{-1}$ & Min & Max & X-points & total \\
			\hline
			206.25                  & 110 & 116 & 226     & 452   \\
			250.00                  & 144 & 165 & 309     & 618   \\
			292.50                  & 159 & 159 & 318     & 636   \\
		\end{tabular}
	\end{ruledtabular}
\end{table}

Across the three times we find 853 critical saddle points. The first snapshot
of the three is shown in Fig. \ref{big_psi}. The first panel shows
the whole domain and the X-points identified within it. 
At first inspection there are only a handful of locations that resemble the classical picture of reconnection synonymous with PIC reconnection simulations (long straight oppositely directed field lines with large coherent current sheets e.g. $x = 77\, d_i$, $y = 22\, d_i$). There are, however, numerous regions with many X-points clustered together. These frequently 
corresponds to ``secondary islands''. 
The second panel shows the zoomed-in region denoted by the black square 
in the first panel, and it becomes clear that the X-points marked in the 
simulation do in fact correspond to critical points of the magnetic vector potential

\begin{figure}
	\includegraphics[width=3.2in]{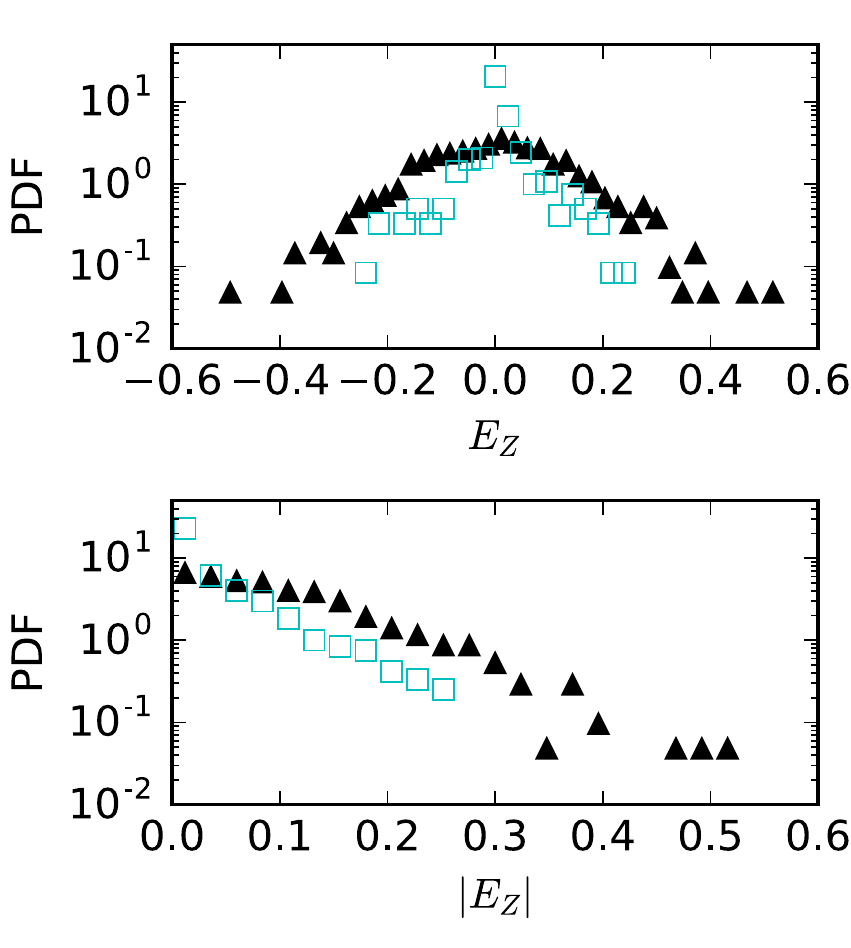}
	\caption{PDF of the reconnection rates. (a) $E_z$ and (b) $|E_z|$ at the X-points identified in the three different times of the large PIC simulation (black triangles) and the values found in a $2048^2$ MHD turbulence simulation (Run 6 in \citet{Wan10}, cyan squares).}
\label{rrates}
\end{figure}

Because of the 2D nature of this simulation we know that reconnection must
occur in the X-Y plane, and so the reconnecting electric field must
point out-of-plane (Z direction). So for each identified saddle point we
interpolate the Fourier filtered, out-of-plane electric field. We generate
the PDFs for $E_{z}$ and $|E_{z}|$ shown as the black triangles in the two
panels of Fig~\ref{rrates}. 
We normalize $E_z$ to the root mean square of the in plane magnetic field (note, in this simulation $B_{x,y,rms} \approx 1$) and so Fig~\ref{rrates} can be interpreted as the PDF of reconnection rates in our simulation. From this, we find the reconnection rates in our kinetic simulation can be as large as .5, with an average magnitude of about 0.1. Fig \ref{rrates} also includes cyan squares that represent the reconnection rates found using the same procedure applied to MHD. This data is from Fig~10 in \citet{WanPP10}. While the MHD and the PIC results have some similarities, it is clear that the PIC distribution is broader than the MHD distribution, in both the range of reconnection rates and the shape of the PDF. It is also worth noting that the average magnitude  in the MHD case ($0.044$) is a little more than a factor of 2 less than the PIC case ($0.10$). This result reaffirms the idea that turbulence can potentially lead to enhanced reconnection rates and is in apparent agreement with the idea that reconnection rates in kinetic plasmas tend to be larger than in MHD and of order 0.1 \cite{Shay99}.

\section{Conclusions}
In this work we have begun to examine the statistics of x-type critical points (X-points) in fully kinetic 2.5D PIC simulations. This work extends the procedures applied to MHD simulations \cite{ServidioPRL09,ServidioPP10} to PIC. We find that noise fluctuations in the magnetic field associated with the counting statistics corresponding to the number of particles per cell (ppc) result in a noisy magnetic vector potential, and ultimately spurious numbers of X-points. Increasing the number of ppc helps to lessen this effect. Other noise reduction techniques examined include post processing the simulation magnetic field data by using either time averaging over a plasma frequency, or spacial filtering using a Fourier or Gaussian method. We showed how each of these affected different statistics of the turbulence, including the omni-directional energy spectrum, the probability density function, the scale dependent kurtosis (SDK) and the number of X-points. We find that the number of ppc would need to be increased several orders of magnitude to have accurate enough field data to identify the correct number of X-points (a prospect that is currently computationally intractable), while time averaging significantly alters the spectrum of the electric field. However, imposing a spatial filter at the Debye scale stabilizes the number of X-points, regardless of the number of ppc, while not dramatically changing the spectrum or the scale dependent kurtosis. The above tests were carried out using kinetic PIC simulation for both Orszag Tang vortex configurations \cite{ParasharPP09} and 
larger broad band turbulence simulations \cite{Wu13}. 

With these filtering techniques we identify the X-points in a large decaying turbulent PIC simulation, and we calculate the reconnection rate at each of these points. We find that the magnitude of the reconnection rates range between 0 and 0.5 in standard normalized Alfv\'en units 
and have an average value 0.1 which is approximately a factor of 2 larger than the MHD result \cite{WanPP10}. Note that reconnection rates of this magnitude are ordinarily associated with a ``fast reconnection'' process 
although in the  present case the normalization is with respect to the global r.m.s fields rather than the local upstream quantities. The PDF of the PIC reconnection rates has a broader shape than in MHD which implies that there are a larger fraction of X-points that have large reconnection rates.
This is consistent with idea that reconnection rates can be boosted by kinetic plasma processes\cite{Shay99}. 

While this work demonstrates a clear procedure to identify X-points in a kinetic PIC simulation, it does not address questions about how varying the number of particle per cell affects the proliferation of X-points during a simulation. It is clear in this work that the number of ppc is an important quantity for the accuracy of the fields, and it has been shown in the MHD case that spatial resolution has a dramatic effect on the number of X-points generated during a simulation \cite{WanPP10}. It is clearly plausible that PIC simulations could be susceptible to a similar issue related to poor counting statistics and spatial resolution. This is an important question for the PIC modeling community that should be addressed in greater detail in the future. In addition, there remain  important questions regarding the {\it physical}, rather than numerical proliferation of X-points in turbulent plasma. This phenomenon has been previously demonstrated for MHD \cite{WanPP13}, where for large systems, at high magnetic Reynold number, the expected number of islands can increase dramatically, even when numerical inaccuracies are carefully controlled. The turbulent proliferation of reconnection sites is clearly a nonlinear dynamical effect, but is likely related to the family of linear instabilities known as plasmoid instabilities \cite{Loureiro07,PucciVellApJi14} that are initiated from equilibrium field configurations. Accurately tracking the putative physical proliferation of X-points in kinetic turbulence would require simultaneously computing the dynamics of a large, high effective Reynolds number plasma, while  respecting the numerical issues we have explored in the present analysis. In this regard the present study is only an initial step in trying to answer these larger questions regarding reconnection in kinetic plasma turbulence.

\begin{acknowledgments}
Research is supported by NSF AGS-1460130
(SHINE), and NASA grant NNX15AW58G, NNX14AI63G (Heliophysics Grand challenge Theory), NNX08A083G–MMS,
the MMS mission through grant NNX14AC39G,
and the Solar Probe Plus science team (IOSIS/Princeton subcontract No. D99031L). Simulations and analysis were performed at NCAR-CISL and at NERSC. The data used are listed in the text, references, and are available by request.

\end{acknowledgments}

\bibliography{parashar-ref,haggerty-ref}

\end{document}